\documentclass[twocolumn,showpacs,prb,superscriptaddress,amsmath,amssymb,floatfix,eqsecnum]{revtex4}
\usepackage{amsmath,amssymb}
\usepackage{bm}
\usepackage{epsfig}
\usepackage{psfrag}

\newcommand{\bd}{\bm}

\begin{document}

\title{Ward identities for the Anderson impurity model: 
derivation via functional methods and the exact renormalization group}

\author{Peter Kopietz}
\affiliation{Institut f\"{u}r Theoretische Physik, Universit\"{a}t
  Frankfurt,  Max-von-Laue Strasse 1, 60438 Frankfurt, Germany}

\author{Lorenz Bartosch}
\affiliation{Institut f\"{u}r Theoretische Physik, Universit\"{a}t
  Frankfurt,  Max-von-Laue Strasse 1, 60438 Frankfurt, Germany}

\author{Lucio Costa}
\affiliation{Institut f\"{u}r Theoretische Physik, Universit\"{a}t
  Frankfurt,  Max-von-Laue Strasse 1, 60438 Frankfurt, Germany}
\affiliation{Centro de Ci\^{e}ncias Naturais e Humanas,
Universidade Federal do ABC,  09210-170 Santo Andr\'{e}, Brazil}

\author{Aldo Isidori}
\affiliation{Institut f\"{u}r Theoretische Physik, Universit\"{a}t
  Frankfurt,  Max-von-Laue Strasse 1, 60438 Frankfurt, Germany}

\author{Alvaro Ferraz}
\affiliation{International Institute for Physics, 
 Universidade Federal do  Rio Grande do Norte, 
59012-970 Natal-RN, Brazil}

\date{March 9, 2010}

 \begin{abstract}

Using functional methods  and the exact renormalization
group we derive Ward identities 
for the Anderson impurity model.
In particular, 
we present a non-perturbative proof  of the 
Yamada-Yosida identities relating
certain coefficients in the low-energy expansion of the self-energy to 
thermodynamic  particle number and spin  susceptibilities of the impurity.
Our proof underlines  the relation of the 
Yamada-Yosida identities to 
the $U(1) \times U(1)$ symmetry
associated with particle number and spin conservation 
in a magnetic field.
\end{abstract}

\pacs{72.15.Qm,71.27.+q,71.10.Pm}

\maketitle

\section{Introduction}

In quantum field theory
symmetries and the associated conservation laws
imply Ward identities, which are exact relations between different types
of Green functions or vertex functions.\cite{ZinnJustin02}
The  constraints imposed by  Ward identities
can be very useful to devise accurate approximation schemes which
do not violate conservation laws.
In this work we shall present a non-perturbative derivation
of the Ward identities relating the coefficients in the low-frequency
expansion of the retarded self-energy $\Sigma ( \omega + i 0 )$
of the Anderson impurity model (AIM) to certain thermodynamic susceptibilities.
A perturbative derivation of these identities has first been given by  
Yamada and Yosida.\cite{Yamada75} 
For particle-hole symmetric filling and in the limit of an infinite bandwidth 
of the conduction electron dispersion with flat density of states these identities read
 \begin{eqnarray}
  {\rm Re}\, \Sigma ( \omega + i 0 )  & = &  \frac{U}{2} + 
\left( 1- \frac{ \tilde{\chi}_{c} + \tilde{\chi}_s }{2} \right)  \omega + {\cal{O}} ( \omega^2 ),
 \hspace{7mm}
 \label{eq:ResigmaWard}
  \\
 {\rm Im}\, \Sigma ( \omega + i 0 ) & = &
  -   \left( \frac{\tilde{\chi}_{c} - \tilde{\chi}_s}{2} \right)^2  \frac{\omega^2}{2 \Delta}
+ {\cal{O}} ( \omega^3 ).
 \label{eq:ImsigmaWard}
\end{eqnarray}
Here, $U$ is the on-site interaction at the impurity site,
$\Delta$ is the imaginary part of the hybridization function
in the limit of an infinite bandwidth of the conduction electron band, 
and $\tilde{\chi}_{c}$ and $\tilde{\chi}_s$ are dimensionless 
particle number (charge) and spin susceptibilities, which will be defined 
in Eqs.~(\ref{eq:chic}) and (\ref{eq:chis}) below.
A generalization 
of Eqs.~(\ref{eq:ResigmaWard}) and (\ref{eq:ImsigmaWard})
to the AIM out of equilibrium can be found in Ref.~[\onlinecite{Oguri01}].

Yamada and Yosida obtained the above identities 
by comparing the coefficients in the perturbation series 
of both sides of Eqs.~(\ref{eq:ResigmaWard}) and (\ref{eq:ImsigmaWard})
to all orders in powers of $U / \Delta$.
An alternative derivation using diagrammatic techniques
can be found in the book by Hewson.\cite{Hewson93}
Unfortunately, in both approaches 
the close relation of the above identities to
the $U(1) \times U(1)$ symmetry associated with
particle number and spin conservation
of the AIM in a magnetic field
is not manifest.
Although powerful functional methods
for deriving Ward identities non-perturbatively 
are well known, \cite{ZinnJustin02} 
apparently there exists no derivation of
Eqs.~(\ref{eq:ResigmaWard}) and (\ref{eq:ImsigmaWard}) in the literature
using these functional methods.
In this work we shall present such a non-perturbative proof of
Eqs.~(\ref{eq:ResigmaWard}) and (\ref{eq:ImsigmaWard})  by combining
standard functional techniques \cite{ZinnJustin02}
with certain exact relations between derivatives of the
self-energy with respect to frequency, chemical potential and magnetic field
which we derive within the framework of the
exact renormalization group.\cite{Berges02,Kopietz10}

To set the stage for our calculation and to define our notation, let us recall  that
the single-site Anderson impurity model is defined in terms of the following second
quantized Hamiltonian,
\begin{eqnarray}
 \hat{H} & = & \sum_{ \bd{k} \sigma} ( \epsilon_{ \bd{k} } - \sigma h )
 \hat{c}^{\dagger}_{ \bd{k} \sigma } \hat{c}_{ \bd{k} \sigma } 
 \nonumber  
\\
 & + & 
  \sum_{\sigma} ( E_d - \sigma h) \hat{d}^{\dagger}_{\sigma} \hat{d}_{\sigma} + U 
    \hat{d}^{\dagger}_{\uparrow} \hat{d}_{\uparrow}
  \hat{d}^{\dagger}_{\downarrow} \hat{d}_{\downarrow}
 \nonumber
 \\
 & + & \sum_{\bd{k} \sigma} ( V_{\bd{k}}^{\ast} 
 \hat{d}^{\dagger}_{ \sigma} \hat{c}_{\bd{k} \sigma } + V_{\bd{k}} 
   \hat{c}^{\dagger}_{\bd{k} \sigma } \hat{d}_{\sigma} ),
 \label{eq:Hdef}
 \end{eqnarray}
where
$\hat{c}_{\bd{k} \sigma}$ annihilates a non-interacting conduction electron with
momentum ${\bd{k}}$, energy dispersion $\epsilon_{\bd{k}}$
and spin projection $\sigma$,
while the operator $\hat{d}_{\sigma}$ annihilates
 a localized correlated $d$-electron with
atomic energy $E_d$ and spin projection $\sigma$.
The hybridization between 
the $d$-electrons and the conduction electrons is characterized by the
hybridization energy $V_{\bd{k}}$, and $h$ is 
 the Zeemann energy
associated with an external magnetic field.
Since we are only interested in the correlation functions of the impurity,
we integrate over the conduction electrons
using the functional integral
representation of the model.\cite{Negele88}
The generating functional
${\cal{G}}_c [ \bar{\jmath}_{\sigma} , j_{\sigma}  ]$ of the connected Green functions
can then be represented as the following ratio of
fermionic functional integrals,
 \begin{equation}
 e^{ {\cal{G}}_c [  \bar{\jmath}_{\sigma} , j_{\sigma} ]  } =
 \frac{ \int {\cal{D}} [ {d} , \bar{d} ] e^{ - S [ \bar{d}_{\sigma} , d_{\sigma} ]  + (  \bar{\jmath} , d)  +  (  \bar{d} , j )    }}{  \int {\cal{D}} [ {d} , \bar{d} ] e^{ - S_0 [ \bar{d}_{\sigma} , 
d_{\sigma} ]  } }
 ,
 \label{eq:Gratio}
 \end{equation}
with the Euclidean action given by
 \begin{eqnarray}
S [ \bar{d}_{\sigma} , d_{\sigma} ] & = & S_0 [ \bar{d}_{\sigma} , d_{\sigma} ]  +
 S_U [\bar{d}_{\sigma} , {d}_{\sigma} ]
  \nonumber
 \\
&=  &
    - \int_{ \omega} \sum_{  \sigma  } G_{ 0, \sigma}^{-1} ( i \omega )
  \bar{d}_{  \omega    \sigma}   d_{  \omega  \sigma}
 \nonumber
 \\
& + &  
U \int_0^{\beta} d \tau \,
\bar{d}_{\uparrow} ( \tau ) d_{\uparrow} ( \tau )
 \bar{d}_{\downarrow} ( \tau ) d_{\downarrow} ( \tau ) .
 \hspace{7mm}
 \label{eq:SSSdef}
 \end{eqnarray}
Here, $\int_{\omega} = \frac{1}{\beta} \sum_{ \omega} $
denotes  summation over fermionic Matsubara frequencies $ i \omega$ and 
$\int_{0}^{\beta} d \tau$ denotes integration over imaginary time,
where $\beta$ is the inverse temperature.
The non-interacting Green function is
 \begin{equation}
 G_{ 0, \sigma} ( i \omega ) = \frac{1}{ i \omega - \xi_{0, \sigma} - \Delta_{\sigma} ( i \omega ) },
 \end{equation}
where 
 \begin{equation}
\xi_{0 , \sigma} = E_d - \mu - \sigma h
 \end{equation}
is the energy of a localized $d$-electron with spin projection $\sigma$
relative to the chemical potential $\mu$, 
and the spin-dependent hybridization function is given by
 \begin{equation}
 \Delta_{\sigma} ( i \omega ) = \sum_{\bd{k}} \frac{ | V_{\bd{k}} |^2 }{ i \omega
 - \epsilon_{\bd{k}} + \mu + \sigma h }.
 \label{eq:Deltadef}
 \end{equation}
The Fourier transform of the Grassmann fields 
$d_{ \sigma} ( \tau )$ in frequency space is defined by
 \begin{equation}
 d_{\sigma} ( \tau ) = \int_{\omega}  e^{ - i \omega \tau } d_{  \omega  \sigma}
 .
\end{equation}
The functional ${\cal{G}}_c [  \bar{\jmath}_{\sigma} , j_{\sigma} ] $
in Eq.~(\ref{eq:Gratio}) depends on Grassmann sources
$\bar{\jmath}_{\sigma}$ and $j_{\sigma}$ and
we have introduced the following notation for the source terms,
 \begin{equation}
  (  \bar{\jmath} , d) = \int_{\omega} \sum_\sigma \bar{\jmath}_{\omega \sigma}
 d_{\omega \sigma}  \; \; , \; \; 
 (  \bar{d} , j) = \int_{\omega} \sum_\sigma \bar{d}_{\omega \sigma}
 j_{\omega \sigma} .
 \end{equation}

\section{Functional Ward identities}

\subsection{$U(1)$ Ward identities due to particle number and spin conservation in a magnetic field}
\label{subsec:U1Ward}

The Euclidean action for the correlated impurity 
given in Eq.~(\ref{eq:SSSdef})  is invariant under
independent global $U(1)$ transformations of the fields for a given spin projection.
This symmetry implies that the generating functional
${\cal{G}}_c [  \bar{\jmath}_{\sigma} , j_{\sigma} ] $ satisfies
certain functional differential equations, so-called functional Ward identities.
By taking functional derivatives of these relations, we shall derive 
the Yamada-Yosida  identities for the self-energy.
Following Refs.~[\onlinecite{ZinnJustin02,Kopietz10}],
we perform a local gauge transformation
on the fermion fields in the (imaginary) time domain,
 \begin{equation}
 d_{\sigma} ( \tau ) =  e^{ - i \alpha_{\sigma} ( \tau ) } d^{\prime}_{\sigma} ( \tau )  
 \; \; , \; \; 
 \bar{d}_{\sigma} ( \tau )  =  e^{  i \alpha_{\sigma} ( \tau ) } 
\bar{d}^{\prime}_{\sigma} ( \tau ) ,
\label{eq:gauge2} 
\end{equation}
where $\alpha_{\sigma} ( \tau )$ are arbitrary real functions.
 The interaction part $S_U$ of our action $S  = S_0   + S_U$
is  invariant under these transformations,  so that
 \begin{eqnarray}
 & & S [ e^{  i \alpha_{\sigma} } \bar{d}_{\sigma}^{\prime} ,
e^{ - i \alpha_{\sigma} } {d}_{\sigma}^{\prime}  ]  =   S [ \bar{d}_{\sigma}^{\prime} ,
{d}_{\sigma}^{\prime} ]
 \nonumber
 \\
& - &     i \int_0^{\beta} d \tau \sum_{\sigma}
\bar{d}_{\sigma}^{\prime} ( \tau ) [ \partial_{\tau} \alpha_{\sigma} ( \tau ) ]
 d^{\prime}_{\sigma} ( \tau )  
 \nonumber
 \\
& + &  i \int_0^{\beta} d \tau  \int_0^{\beta}    d \tau^{\prime} \sum_{\sigma}
\bar{d}_{\sigma}^{\prime} ( \tau )  [  \alpha_{\sigma} ( \tau )
- \alpha_{\sigma} ( \tau^{\prime}  ) ] 
 \nonumber
 \\
& & 
\times \Delta_{\sigma} ( \tau - \tau^{\prime} ) 
 d^{\prime}_{\sigma} ( \tau^{\prime} )  + {\cal{O}} \left( \alpha^2  \right),
 \label{eq:Sinvariance}
\end{eqnarray}
where
 $\Delta_{\sigma} ( \tau ) = \int_{\omega}  e^{ - i \omega \tau } 
\Delta_{\sigma} ( i \omega )$.
Using the invariance of the functional integral representation (\ref{eq:Gratio}) of
${\cal{G}}_c [ \bar{\jmath}_{\sigma} , j_{\sigma}  ]$
with respect to a change of the integration  variables
$d_{\sigma} \rightarrow d_{\sigma}^{\prime}$, 
$\bar{d}_{\sigma} \rightarrow \bar{d}_{\sigma}^{\prime}$, 
 and expanding to
linear order in the gauge factors $\alpha_{\sigma} ( \tau )$ we obtain the
desired functional Ward identity. \cite{ZinnJustin02,Kopietz10}
For our purpose it is convenient to express 
the ``current terms'' of this Ward identity via the
generating functional $\Gamma [ \bar{d}_{\sigma} , d_{\sigma} ]$ of the
irreducible vertices, which is obtained from 
${\cal{G}}_c [ \bar{\jmath}_{\sigma} , j_{\sigma}  ]$ via a functional
Legendre transformation,~\cite{ZinnJustin02,Kopietz10}
 \begin{equation}
\Gamma [ \bar{d}_{\sigma} , d_{\sigma} ] =  
(  \bar{\jmath} , d)  +  (  \bar{d} , j) - {\cal{G}}_c [ \bar{\jmath}_{\sigma} , j_{\sigma} ]
-  S_0[ \bar{d}_{\sigma} , d_{\sigma} ],
\end{equation}
where on the right-hand side it is understood that the sources
$\bar{\jmath}_{\sigma}$ and $j_{\sigma}$
should be calculated as functions of the
field averages $\bar{d}_{\sigma}$ and $d_{\sigma}$ by solving the
equations
 \begin{equation}
 d_{\sigma} = \frac{ \delta {\cal{G}}_c   [ \bar{\jmath}_{\sigma} , j_{\sigma} ]}{
 \delta \bar{\jmath}_{\sigma} } \; \; , \; \;
\bar{d}_{\sigma} = - \frac{ \delta {\cal{G}}_c   [ \bar{\jmath}_{\sigma} , j_{\sigma} ]}{
 \delta j_{\sigma} }\; .
 \end{equation}
After Fourier transformation to frequency space and some 
rearrangements analogous  to those in Refs.~[\onlinecite{Kopietz10,Schuetz05}] we obtain the functional Ward identity
\begin{eqnarray}
& &  \int_{ \omega^{\prime}}    \biggl\{  \left[   
G_{0, \sigma^{\prime}}^{-1} ( i \omega^{\prime} + i \bar{\omega} )   
- G_{0, \sigma^{\prime}}^{-1} ( i \omega^{\prime} ) 
 \right]
 \frac{ \delta^2 {\cal{G}}_c }{ \delta \bar{\jmath}_{ \omega^{\prime} \sigma^{\prime}}
 \delta j_{ \omega^{\prime} + \bar{\omega}, \sigma^{\prime} } }
 \nonumber
 \\
 & & 
  \hspace{9mm} + 
 d_{ \omega^{\prime} \sigma^{\prime}}
  \frac{ \delta \Gamma }{ \delta d_{ \omega^{\prime} + \bar{\omega}, \sigma^{\prime}} }
  - 
\bar{d}_{ \omega^{\prime} + \bar{\omega}, \sigma^{\prime} }
  \frac{ \delta \Gamma }{ \delta \bar{d}_{ \omega^{\prime}  \sigma^{\prime}}} 
  \biggr\} 
= 0 ,
 \label{eq:Wardfermi}
 \end{eqnarray}
where $\bar{\omega}$ is an external bosonic Matsubara frequency.
Summing both sides of this functional equation over $\sigma^{\prime}$ 
we obtain the functional Ward identity due to the $U(1)$ symmetry associated with
particle number conservation,
\begin{eqnarray}
& &  \hspace{-3mm} \int_{ \omega^{\prime}} \sum_{\sigma^{\prime}}   \biggl\{  \left[   
G_{0, \sigma^{\prime}}^{-1} ( i \omega^{\prime} + i \bar{\omega} )   - 
G_{0, \sigma^{\prime}}^{-1} ( i \omega^{\prime} ) 
 \right]
 \frac{ \delta^2 {\cal{G}}_c }{ \delta \bar{\jmath}_{ \omega^{\prime} \sigma^{\prime}}
 \delta j_{ \omega^{\prime} + \bar{\omega}, \sigma^{\prime} } }
 \nonumber
 \\
 & & 
  \hspace{9mm} + 
 d_{ \omega^{\prime} \sigma^{\prime}}
  \frac{ \delta \Gamma }{ \delta d_{ \omega^{\prime} + \bar{\omega}, \sigma^{\prime}} }
  - 
\bar{d}_{ \omega^{\prime} + \bar{\omega}, \sigma^{\prime} }
  \frac{ \delta \Gamma }{ \delta \bar{d}_{ \omega^{\prime}  \sigma^{\prime}}} 
  \biggr\} 
= 0 \; .
 \label{eq:WardU1charge}
 \end{eqnarray}
To obtain the analogous $U(1)$ Ward identity associated with conservation of the
spin projection along the axis of the magnetic field,
we multiply Eq.~(\ref{eq:Wardfermi}) by $\sigma^{\prime}$ before summing
over $\sigma^{\prime}$, which yields
\begin{eqnarray}
& & \hspace{-6mm}  \int_{ \omega^{\prime}} \sum_{\sigma^{\prime}}  \sigma^{\prime} \biggl\{  \left[   
G_{0, \sigma^{\prime}}^{-1} ( i \omega^{\prime} + i \bar{\omega} )   - 
G_{0, \sigma^{\prime}}^{-1} ( i \omega^{\prime} ) 
 \right]
 \frac{ \delta^2 {\cal{G}}_c }{ \delta \bar{\jmath}_{ \omega^{\prime} \sigma^{\prime}}
 \delta j_{ \omega^{\prime} + \bar{\omega}, \sigma^{\prime} } }
 \nonumber
 \\
 & & 
  \hspace{9mm} + 
 d_{ \omega^{\prime} \sigma^{\prime}}
  \frac{ \delta \Gamma }{ \delta d_{ \omega^{\prime} + \bar{\omega}, \sigma^{\prime}} }
  - 
\bar{d}_{ \omega^{\prime} + \bar{\omega}, \sigma^{\prime} }
  \frac{ \delta \Gamma }{ \delta \bar{d}_{ \omega^{\prime}  \sigma^{\prime}}} 
  \biggr\} 
= 0 \; .
 \label{eq:WardU1spin}
 \end{eqnarray}

\subsection{$SU(2)$ Ward identity due to spin conservation}
\label{subsec:SU2}

In the absence of a magnetic field our action (\ref{eq:SSSdef})
is invariant under arbitrary rotations in spin space.
To derive the corresponding $SU(2)$ Ward identity,
we perform a local rotation in spin space,
 \begin{equation}
\left( \begin{array}{c} d_{\uparrow} ( \tau ) \\
 d_{\downarrow} ( \tau ) \end{array} \right)
 = U ( \tau ) \left( \begin{array}{c} d^{\prime}_{\uparrow} ( \tau ) \\
 d_{\downarrow}^{\prime} ( \tau ) \end{array} \right),
 \label{eq:SU2gauge} 
\end{equation}
where the $SU(2)$ matrix $U ( \tau )$ can be written as
 \begin{equation}
 U ( \tau )  =  
e^{ - i  \bd{\sigma} \cdot \bd{\alpha} ( \tau )}.
 \label{eq:SU2Udef}
\end{equation}
Here, $ \bd{\sigma} = [ \sigma^x , \sigma^y , \sigma^z ]$ 
is the vector of Pauli matrices and
$\bd{\alpha} ( \tau )$ is a time-dependent three-component vector.
Expanding to linear order in $ \bd{\alpha} ( \tau )$
we obtain, after the same manipulations as in
Sec.~\ref{subsec:U1Ward}, the following $SU(2)$ Ward identity,
\begin{eqnarray}
& & \hspace{-4mm}  \int_{ \omega^{\prime}} \sum_{ \sigma \sigma^{\prime}} 
{\sigma}^{i}_{ \sigma \sigma^{\prime}}
  \biggl\{  \left[   
G_0^{-1} ( i \omega^{\prime} + i \bar{\omega} )   - G_0^{-1} ( i \omega^{\prime} ) 
 \right]
\frac{ \delta^2 {\cal{G}}_c }{ \delta \bar{\jmath}_{ \omega^{\prime} \sigma^{\prime}}
 \delta j_{ \omega^{\prime} + \bar{\omega}, \sigma } }
 \nonumber
 \\
 & &   
 \hspace{4mm} {} +
 d_{ \omega^{\prime} \sigma^{\prime}}
  \frac{ \delta \Gamma }{ \delta d_{ \omega^{\prime} + \bar{\omega}, \sigma} }
  - 
\bar{d}_{ \omega^{\prime} + \bar{\omega}, \sigma }
  \frac{ \delta \Gamma }{ \delta \bar{d}_{ \omega^{\prime}  \sigma^{\prime}}} 
  \biggr\}
= 0 ,
 \label{eq:WardSU2spin}
 \end{eqnarray}
where the superscript $i = x,y,z$ labels the three components of the
vector operator $\bd{\sigma}$.
Together with the particle number conservation Ward identity (\ref{eq:WardU1charge})
these equations are equivalent to the four Ward identities 
\begin{eqnarray}
& & \hspace{-4mm}  \int_{ \omega^{\prime}} 
  \biggl\{  \left[   
G_0^{-1} ( i \omega^{\prime} + i \bar{\omega} )   - G_0^{-1} ( i \omega^{\prime} ) 
 \right]
\frac{ \delta^2 {\cal{G}}_c }{ \delta \bar{\jmath}_{ \omega^{\prime} \sigma^{\prime}}
 \delta j_{ \omega^{\prime} + \bar{\omega}, \sigma } }
 \nonumber
 \\
 & &   
 \hspace{2mm} {} +
 d_{ \omega^{\prime} \sigma^{\prime}}
  \frac{ \delta \Gamma }{ \delta d_{ \omega^{\prime} + \bar{\omega}, \sigma} }
  - 
\bar{d}_{ \omega^{\prime} + \bar{\omega}, \sigma }
  \frac{ \delta \Gamma }{ \delta \bar{d}_{ \omega^{\prime}  \sigma^{\prime}}} 
  \biggr\}
= 0 ,
 \label{eq:WardSU2spinsingle}
 \end{eqnarray}
where $\sigma , \sigma^{\prime} \in \{ \uparrow, \downarrow \}$  
are now fixed spin projections.
Note that in the absence of a magnetic field the 
Green function and the self-energy are independent of the spin quantum number
$\sigma$,  so that we may write
$G_{\sigma} ( i \omega ) = G ( i \omega )$ and $\Sigma_{\sigma} ( i \omega ) =
\Sigma ( i \omega )$.
For $i = z$ we have ${\sigma}^{z}_{ \sigma \sigma^{\prime}} =
\sigma^{\prime} \delta_{ \sigma , \sigma^{\prime}}$ so that
Eq.~(\ref{eq:WardSU2spin}) reduces to the zero-field limit of the
$U(1)$ Ward identity (\ref{eq:WardU1spin}) associated with the
conservation of the spin component in the direction of the magnetic
field.

\section{Ward identities for the self-energy}

\subsection{Particle number conservation}

For the AIM in a magnetic field, the first terms in the functional
Taylor expansion of the generating functional
$\Gamma [ \bar{d}_{\sigma} , d_{\sigma} ]$ are of the form \cite{Kopietz10}
 \begin{eqnarray}
 \Gamma [ \bar{d}_{\sigma} , d_{\sigma} ] & = & \Gamma_0 + \int_{\omega}
 \sum_{\sigma} \Sigma_{\sigma} ( i \omega ) \bar{d}_{ \omega \sigma} 
d_{\omega \sigma} \nonumber
 \\
 &    & \hspace{-23mm}  +\frac{1}{2}
  \int_{ \omega_1^{\prime}}\int_{ \omega_2^{\prime}}
 \int_{ \omega_2}\int_{ \omega_1} \sum_{ \sigma \sigma^{\prime}} 
 \beta \delta_{ \omega_1^{\prime} + \omega_2^{\prime} , \omega_2 + \omega_1 }
 \nonumber
 \\
 & & \hspace{-20mm} \times
U^{(4)}_{ \sigma , \sigma^{\prime}} ( i \omega_1^{\prime} , i \omega_2^{\prime} ;
 i \omega_2 , i \omega_1 ) 
\bar{d}_{ \omega_1^{\prime} \sigma}
\bar{d}_{ \omega_2^{\prime} \sigma^{\prime}} d_{\omega_2 \sigma^{\prime} }
 d_{\omega_1 \sigma} 
 \nonumber
 \\
 & &   \hspace{-23mm}  + {\cal{O}} \left( \bar{d}^3  d^3 \right),
 \end{eqnarray}
where $\Gamma_0$ is proportional to the interaction correction
to the grand canonical potential, $\Sigma_{\sigma} ( i \omega )$ is the exact irreducible
self-energy, and $U^{(4)}_{ \sigma , \sigma^{\prime}} ( i \omega_1^{\prime} , i \omega_2^{\prime} ;
 i \omega_2 , i \omega_1 ) $ is the exact effective interaction.
The term in the second line of our functional Ward identity (\ref{eq:WardU1charge}) then
yields
 \begin{eqnarray}
  & & \int_{ \omega^{\prime}} \sum_{\sigma^{\prime}}   \biggr[
 d_{ \omega^{\prime} \sigma^{\prime}}
  \frac{ \delta \Gamma }{ \delta d_{ \omega^{\prime} + \bar{\omega}, \sigma^{\prime}} }
  - 
\bar{d}_{ \omega^{\prime} + \bar{\omega}, \sigma^{\prime} }
  \frac{ \delta \Gamma }{ \delta \bar{d}_{ \omega^{\prime}  \sigma^{\prime}}} 
  \biggr] 
 \nonumber
\\
 &=& 
\int_{ \omega^{\prime}} \sum_{\sigma^{\prime}} 
   [  \Sigma_{\sigma^{\prime}} ( 
i \omega^{\prime} + i \bar{\omega} ) -
 \Sigma_{\sigma^{\prime}} ( i \omega^{\prime} ) ] 
\bar{d}_{ \omega^{\prime} + \bar{\omega} \sigma^{\prime}}
 d_{\omega^{\prime} \sigma^{\prime}}
 \nonumber
 \\
 & & 
+  {\cal{O}} \left( \bar{d}^2  d^2  \right). 
 \label{eq:WardUchargeL2}
 \end{eqnarray}
To extract the Ward identity for the self-energy from
our functional Ward identity  (\ref{eq:WardU1charge}),
we take the second functional derivative
 $
 \frac{\delta}{\delta d_{ \omega \sigma}} 
 \frac{\delta}{\delta \bar{d}_{ \omega + \bar{\omega}, \sigma}} 
 $
of both sides of Eq.~(\ref{eq:WardU1charge})
and set then all fields equal to zero, which yields
 \begin{eqnarray}
 & & \Sigma_{\sigma} ( i \omega + i \bar{\omega} ) - \Sigma_{\sigma} ( i \omega ) 
 \nonumber
 \\
& = & 
 - \int_{\omega^{\prime}}  \sum_{\sigma^{\prime}}
\left[   
G_{0, \sigma^{\prime}} ^{-1} ( i \omega^{\prime} + i \bar{\omega} )   
- G_{0, \sigma^{\prime}}^{-1} ( i \omega^{\prime} ) 
 \right] 
 \nonumber
 \\
 & &  \hspace{7mm} \times
G_{\sigma^{\prime}} ( i \omega^{\prime} + i \bar{\omega} ) 
G_{\sigma^{\prime}} ( i \omega^{\prime} )
 \nonumber
 \\
 & & \hspace{7mm} \times
U^{(4)}_{\sigma , \sigma^{\prime}} ( i \omega + i \bar{\omega}  ,
 i \omega^{\prime} ; i \omega^{\prime} + i \bar{\omega}  ,
i \omega  ).
 \label{eq:Wardparticle}
 \end{eqnarray}
To obtain the right-hand side of this Ward identity, we have used 
 \begin{eqnarray}
  & &\left.
\frac{\delta}{\delta d_{ \omega \sigma}} 
 \frac{\delta}{\delta \bar{d}_{ \omega + \bar{\omega}, \sigma}} 
  \frac{ \delta^2 {\cal{G}}_c }{ \delta \bar{\jmath}_{ \omega^{\prime} \sigma^{\prime}}
 \delta j_{ \omega^{\prime} + \bar{\omega}, \sigma^{\prime} } } \right|_{ {\rm fields} =0 }
 \nonumber 
\\
 &=  &  
G_{\sigma^{\prime}} ( i \omega^{\prime} ) G_{\sigma^{\prime}}
 ( i \omega^{\prime} + i \bar{\omega} ) 
\nonumber
 \\
 & & \times
U^{(4)}_{ \sigma , \sigma^{\prime}  }
 (   i \omega + i \bar{\omega} ,  i \omega^{\prime}  ; i \omega^{\prime} +
 i \bar{\omega} ,  i \omega   ),
 \label{eq:tree}
 \end{eqnarray}
which follows from
the tree expansion relating
the connected Green functions generated by  ${\cal{G}}_c $ 
to the irreducible vertices.\cite{Kopietz10,Negele88}
In the limit of vanishing bosonic frequency $\bar{\omega} \rightarrow 0$
the Ward identity (\ref{eq:Wardparticle}) reduces to
 \begin{eqnarray}
 \frac{ \partial \Sigma_{\sigma} ( i \omega  ) }{\partial ( i \omega )} &  = & 
 - \int_{\omega^{\prime}}  \sum_{\sigma^{\prime}}
\left[ 1 - \frac{ \partial 
\Delta_{\sigma^{\prime}} ( i \omega^{\prime} )}{\partial ( i \omega^{\prime} )}   
 \right] G_{\sigma^{\prime}}^2 ( i \omega^{\prime} ) 
 \nonumber
 \\
 &  & \times  \Gamma_{ \sigma, \sigma^{\prime}} ( i \omega , i \omega^{\prime} ),
 \label{eq:WardU1final}
 \end{eqnarray}
where we have defined
 \begin{equation}
 \Gamma_{ \sigma, \sigma^{\prime}} ( i \omega , i \omega^{\prime} ) =
 U^{(4)}_{ \sigma , \sigma^{\prime}  }
 (   i \omega  ,  i \omega^{\prime}  ; i \omega^{\prime}  ,  i \omega   ).
 \label{eq:Gammashort}
 \end{equation}

The above identities are valid for arbitrary hybridization functions $\Delta_{\sigma} ( i \omega )$.
Of special interest is the limit of infinite bandwidth of the 
conduction electron band with flat density of states, where the general expression 
for $\Delta_{\sigma} ( i \omega )$ given in Eq.~(\ref{eq:Deltadef})
is given by
 \begin{equation}
 \Delta_{\sigma} ( i \omega ) = - i \Delta {\rm sgn}\, \omega   = -  i \Delta
[ 2 \Theta ( \omega ) -1 ].
 \label{eq:Deltainfinite}
 \end{equation}
Here, the hybridization energy 
$\Delta$ is assumed to be independent of the chemical potential and
the magnetic field.
In this limit
 \begin{equation}
  \frac{ \partial 
\Delta_{\sigma^{\prime}} ( i \omega^{\prime} )}{\partial ( i \omega^{\prime} )}   
 = - 2 \Delta \delta ( \omega^{\prime} ).
 \label{eq:Deltainfinitederiv}
 \end{equation}
The term $(\partial 
\Delta_{\sigma^{\prime}} ( i \omega^{\prime} ) / \partial ( i \omega^{\prime} ))   
  G_{\sigma^{\prime}}^2 ( i \omega^{\prime} ) $ in
Eq.~(\ref{eq:WardU1final}) is then ambiguous because the $\delta$-function
in the first term is multiplied by the sign-function
associated with 
the hybridization function in  $G_{\sigma^{\prime}}^2 ( i \omega^{\prime} )$. 
To properly define this term one should use the 
Morris-Lemma,~\cite{Morris94,Bartosch09}  which states that the
product of the delta-function with an arbitrary function $f ( \Theta (x))$ of the
$\Theta$-function should be defined via
 \begin{equation}
 \delta ( x ) f ( \Theta ( x ) ) = \delta (x ) \int_0^1 dt f ( t ).
 \end{equation}
We conclude
that in the infinite bandwidth limit
 \begin{eqnarray}
 & &  -\frac{ \partial 
\Delta_{\sigma^{\prime}} ( i \omega^{\prime} )}{\partial ( i \omega^{\prime} )}   
  G_{\sigma^{\prime}}^2 ( i \omega^{\prime} )  =  
 2 \Delta \delta ( \omega^{\prime} )  G_{\sigma^{\prime}}^2 ( i \omega^{\prime} )
\nonumber
 \\
 & = &  2 \Delta \delta ( \omega^{\prime} ) \int_0^1 dt 
 \frac{1}{ [ - \xi_{  \sigma^{\prime}}  + i \Delta ( 2 t -1 )  ]^2 }  
 \nonumber
 \\
 & = & \delta ( \omega^{\prime} )   \frac{ 2 \Delta }{ \xi_{\sigma^{\prime}}^2 + \Delta^2 }  
 \equiv
 2 \pi      \delta ( \omega^{\prime} )  \rho_{\sigma^{\prime}} (0 )   ,
 \label{eq:deltaMorris}
\end{eqnarray}
 where $\xi_{\sigma} = \xi_{ 0 , \sigma} + \Sigma_{\sigma} ( i0)$ 
is the true excitation energy of a localized electron with spin projection $\sigma$,
and
 \begin{equation}
 \rho_{\sigma} (0 ) = \frac{\Delta}{\pi}  \frac{1}{ \xi_{\sigma}^2 + \Delta^2 }
 \label{eq:rhodef}
 \end{equation} 
is the spectral density of the localized $d$-electron with spin $\sigma$
at vanishing energy.
Substituting Eq.~(\ref{eq:deltaMorris}) into
(\ref{eq:WardU1final}) we obtain the corresponding Ward identity in the
infinite bandwidth limit,\cite{footnoteMorris}
\begin{eqnarray}
 \frac{ \partial \Sigma_{\sigma} ( i \omega  ) }{\partial ( i \omega )} &  = & 
 - \int_{\omega^{\prime}}  \sum_{\sigma^{\prime}}
G_{\sigma^{\prime}}^2 ( i \omega^{\prime} )  
\Gamma_{ \sigma , \sigma^{\prime}} ( i \omega , i \omega^{\prime} )
 \nonumber
 \\
 & & -  \sum_{\sigma^{\prime}} \rho_{\sigma^{\prime}} ( 0 )
\Gamma_{ \sigma , \sigma^{\prime}} ( i \omega , 0 ).
 \label{eq:WardU1infinite}
\end{eqnarray}
If we (incorrectly) replace
$\rho_{\sigma^{\prime}} ( 0 ) \rightarrow \rho_{\sigma} (0)$
in the second line of Eq.~(\ref{eq:WardU1infinite}), we arrive at
Eq.~(5.71) of Ref.~[\onlinecite{Hewson93}].

\subsection{Spin  conservation}

The Ward identity for the self-energy associated with
the conservation of the spin projection  in the direction of the magnetic field
can be derived analogously from the corresponding functional
Ward identity (\ref{eq:WardU1spin}).
By comparing the functional Ward identity (\ref{eq:WardU1charge}) 
due to particle number conservation with the corresponding $U(1)$ Ward identity
(\ref{eq:WardU1spin}) due to spin conservation, we conclude that
in the spin case the Ward identities for the self-energy
can be obtained from those of the particle number case by simply replacing
 \begin{eqnarray}
 \Sigma_{\sigma} ( i \omega ) & \rightarrow & \sigma \Sigma_{\sigma} ( i \omega ),
 \\
 \sum_{ \sigma^{\prime}} & \rightarrow  & \sum_{ \sigma^{\prime}} \sigma^{\prime}.
 \end{eqnarray}
In particular, the spin-analogue of the
particle number Ward identity (\ref{eq:WardU1final}) is
 \begin{eqnarray}
 \sigma
 \frac{ \partial \Sigma_{\sigma} ( i \omega  ) }{\partial ( i \omega )} &  = & 
 - \int_{\omega^{\prime}}  \sum_{\sigma^{\prime}} \sigma^{\prime}
\left[ 1 - \frac{ \partial 
\Delta_{\sigma^{\prime}} ( i \omega^{\prime} )}{\partial ( i \omega^{\prime} )}   
 \right] G_{\sigma^{\prime}}^2 ( i \omega^{\prime} ) 
 \nonumber
 \\
 &  & \times
\Gamma_{ \sigma , \sigma^{\prime}} ( i \omega , i \omega^{\prime} ).
 \label{eq:WardselfU1spin}
 \end{eqnarray}
Combining this equation with the corresponding equation for the particle number, 
Eq.~(\ref{eq:WardU1final}), we obtain the two Ward identities
 \begin{eqnarray}
 \frac{ \partial \Sigma_{\sigma} ( i \omega  ) }{\partial ( i \omega )} &  = & 
 - \int_{\omega^{\prime}}  
\left[ 1 - \frac{ \partial 
\Delta_{\sigma} ( i \omega^{\prime} )}{\partial ( i \omega^{\prime} )}   
 \right] G_{\sigma}^2 ( i \omega^{\prime} ) 
 \nonumber
 \\
 &  & \times
\Gamma_{ \sigma , \sigma} ( i \omega , i \omega^{\prime} ) , 
 \label{eq:WardselfU1spinspin} \\
 0 &  = & 
 - \int_{\omega^{\prime}}  
\left[ 1 - \frac{ \partial 
\Delta_{-\sigma} ( i \omega^{\prime} )}{\partial ( i \omega^{\prime} )}   
 \right] G_{-\sigma}^2 ( i \omega^{\prime} ) 
 \nonumber
 \\
 &  & \times
\Gamma_{ \sigma , - \sigma} ( i \omega , i \omega^{\prime} ).
 \label{eq:WardselfU1spinminusspin}
 \end{eqnarray}
In the limit of an infinite bandwidth these 
equations can be written in analogy to Eq.~(\ref{eq:WardU1infinite}) as
\begin{eqnarray}
 \frac{ \partial \Sigma_{\sigma} ( i \omega  ) }{\partial ( i \omega )} &  = & 
 - \int_{\omega^{\prime}}  
G_{\sigma}^2 ( i \omega^{\prime} )  
\Gamma_{ \sigma , \sigma} ( i \omega , i \omega^{\prime} )
 \nonumber
 \\
 & & - \rho_{\sigma} ( 0 )
\Gamma_{ \sigma , \sigma} ( i \omega , 0 ) , 
 \label{eq:WardU1infinitespinspin} \\
 0 & = &
 - \int_{\omega^{\prime}}  
G_{- \sigma}^2 ( i \omega^{\prime} )  
\Gamma_{ \sigma , - \sigma} ( i \omega , i \omega^{\prime} )
 \nonumber
 \\
 & & -  \rho_{- \sigma} ( 0 )
\Gamma_{ \sigma , - \sigma} ( i \omega , 0 ).
 \label{eq:WardU1infinitespinminusspin}
\end{eqnarray}
Finally, we note that in the absence of a magnetic field
the $SU(2)$ functional Ward identity (\ref{eq:WardSU2spin}) 
does not imply any further independent relation for the self-energy, 
since in this case we just have $\Sigma_{\sigma} ( i \omega ) = \Sigma ( i \omega )$. However, taking higher order
functional derivatives of Eq. (\ref{eq:WardSU2spin}) we may obtain 
symmetry relations involving higher order vertices, which are beyond the scope of this 
work.

\section{Proof of the Yamada-Yosida identities}

\subsection{Relations between derivatives of the self-energy}

The above Ward identities can now be used to derive exact relations between
the derivatives of the self-energy with respect to frequency, the chemical potential and
the magnetic field.
In  the book by Hewson~\cite{Hewson93} one can find  a derivation of these identities 
using diagrammatic arguments.
Here, we show that the correct relations
can be obtained quite simply within the framework of the exact 
renormalization
group (also known as the functional renormalization group).\cite{Berges02,Kopietz10}
As a special case of the general exact renormalization group equation for the
irreducible self-energy of an interacting Fermi 
system derived in Refs.~[\onlinecite{Kopietz01,Salmhofer01,Kopietz10}]
we find for the self-energy of the AIM 
 \begin{equation}
 \partial_{\Lambda} \Sigma_{\sigma } (i \omega ) =
 \int_{\omega^{\prime}} \sum_{\sigma^{\prime}}
\dot{G}_{\sigma^{\prime} } ( i \omega^{\prime} )  
\Gamma_{ \sigma , \sigma^{\prime}} ( i \omega , i \omega^{\prime} ),
 \label{eq:selfRGflow}
 \end{equation}
where we have used again the notation (\ref{eq:Gammashort})
for the effective interaction, and
where $\Lambda$ is some flow parameter (cutoff) 
appearing in the Gaussian part of the
action. All Green functions and vertices in Eq.~(\ref{eq:selfRGflow})
implicitly depend on the parameter $\Lambda$.
The so-called single-scale propagator is defined by
 \begin{equation}
 \dot{G}_{\sigma } ( i \omega )  = - 
 {G}^2_{\sigma } ( i \omega )  \partial_{\Lambda} 
 {G}^{-1}_{0,\sigma } ( i \omega ).
 \end{equation}
The exact renormalization group flow equation (\ref{eq:selfRGflow}) 
is valid for any choice of the flow parameter $\Lambda$. In particular, we may
choose the chemical potential $\mu$ as a flow parameter.\cite{Sauli06}
Then $\Lambda = \mu$ and hence
 \begin{eqnarray}
 \partial_{\Lambda} 
 {G}^{-1}_{0,\sigma } ( i \omega )
 & = & \frac{\partial}{\partial \mu } [ i \omega - E_d + \mu + \sigma h - 
\Delta_{\sigma} ( i \omega ) ] 
 \nonumber
 \\
 & = & 1 - \frac{ \partial 
\Delta_{\sigma } ( i \omega )}{\partial \mu} , 
\end{eqnarray}
so that the single-scale propagator is simply
 \begin{equation}
\dot{G}_{\sigma } ( i \omega )  =  - 
 {G}^2_{\sigma } ( i \omega )  \left[ 1 - \frac{ \partial 
\Delta_{\sigma} ( i \omega )}{\partial \mu} \right] .
 \end{equation}
In this chemical potential cutoff scheme the
exact renormalization group flow equation (\ref{eq:selfRGflow})
reduces to
 \begin{eqnarray}
 \frac{\partial \Sigma_{\sigma } (i \omega )}{ \partial \mu } 
 &  = & 
 - \int_{\omega^{\prime}}  \sum_{\sigma^{\prime}}
\left[ 1 - \frac{ \partial 
\Delta_{\sigma^{\prime}} ( i \omega^{\prime} )}{\partial \mu}   
 \right] G_{\sigma^{\prime}}^2 ( i \omega^{\prime} ) 
 \nonumber
 \\
 &  & \times \Gamma_{\sigma , \sigma^{\prime}} ( i \omega , i \omega^{\prime}).
 \label{eq:chemflowfinal}
 \end{eqnarray}
To further manipulate this expression, let us note that for $\omega \neq 0$ the integrand of the hybridization function given in Eq.~(\ref{eq:Deltadef}) is non-singular, such that 
\begin{equation}
 \frac{ \partial 
\Delta_{\sigma} ( i \omega )}{\partial (i \omega)} = 
 \frac{ \partial 
\Delta_{\sigma} ( i \omega )}{\partial \mu } 
, \quad \omega \neq 0.
 \label{eq:Deltaomegaderiv}
\end{equation}
For $\omega = 0$, however, we have to be more careful. Defining the spectral density
 \begin{equation}
  g(\epsilon) = \pi \sum_{\bd{k}}  | V_{\bd{k}} |^2 \delta ( \epsilon - \epsilon_{\bd{k}} ),
 \end{equation}
the hybridization function $\Delta_\sigma(i\omega)$ can be rewritten as
 \begin{equation}
 \Delta_{\sigma} ( i \omega ) = \frac{1}{\pi} \int_{- \infty}^{\infty} d\epsilon\, g(\epsilon) \frac{1}{ i \omega
 - \epsilon + \mu + \sigma h }.
 \label{eq:Deltaspec}
\end{equation}
Using the well-known formula
\begin{equation}
  \frac{1}{x \pm i 0^+} = P \frac{1}{x} \mp i \pi \delta(x),
\end{equation}
where $P$ is the principle value, we see that $\Delta_\sigma(z)$ has a branch cut along the real axis. More explicitly, defining 
\begin{equation}
  \Delta = g(\mu + \sigma h),
\end{equation}
we obtain
 \begin{equation}
 \frac{ \partial 
\Delta_{\sigma} ( i \omega )}{\partial (i \omega)} = 
 \frac{ \partial 
\Delta_{\sigma} ( i \omega )}{\partial \mu }  - 2 \Delta \delta ( \omega ) ,
 \label{eq:Deltaomegaderiv2}
 \end{equation}
where the term $-2 \Delta \delta ( \omega )$ arises from the discontinuity across the
branch cut. 
In the limit of a flat band of  infinite width discussed above, $g(\epsilon) = \Delta$ 
for all $\epsilon$ and $ \partial \Delta_{\sigma} ( i \omega ) /\partial \mu  =0$, so that
Eq.~(\ref{eq:Deltaomegaderiv2})
reduces to Eq.~(\ref{eq:Deltainfinitederiv}).
Using the identity (\ref{eq:Deltaomegaderiv2}) we may write 
our exact renormalization group flow equation (\ref{eq:chemflowfinal}) in the form
 \begin{eqnarray}
 \frac{\partial \Sigma_{\sigma } (i \omega )}{ \partial \mu } 
 &  = & 
 - \int_{\omega^{\prime}}  \sum_{\sigma^{\prime}}
\left[ 1 - \frac{ \partial 
\Delta_{\sigma^{\prime}} ( i \omega^{\prime} )}{\partial ( i \omega^{\prime} )}   
 - 2 \Delta \delta ( \omega^{\prime} ) 
 \right] 
 \nonumber
 \\
 &  & \times G_{\sigma^{\prime}}^2 ( i \omega^{\prime} ) 
\Gamma_{\sigma , \sigma^{\prime}} ( i \omega , i \omega^{\prime})
 \nonumber
 \\
 &  & \hspace{-20mm} = 
 - \int_{\omega^{\prime}}  \sum_{\sigma^{\prime}}
\left[ 1 - \frac{ \partial 
\Delta_{\sigma^{\prime}} ( i \omega^{\prime} )}{\partial ( i \omega^{\prime} )}   
 \right]  G_{\sigma^{\prime}}^2 ( i \omega^{\prime} ) 
\Gamma_{\sigma , \sigma^{\prime}} ( i \omega , i \omega^{\prime})
 \nonumber
 \\
 &  & 
+ \sum_{\sigma^{\prime}} \rho_{\sigma^{\prime}} ( 0 )
\Gamma_{\sigma , \sigma^{\prime}} ( i \omega , 0).
 \label{eq:chemflowfinal2}
 \end{eqnarray}
Comparing the right-hand side of this exact relation with
the right-hand side of the particle number Ward identity
(\ref{eq:WardU1final}) we  conclude that for all frequencies
we have the following exact identity,
\begin{equation}  
 \frac{ \partial \Sigma_{\sigma} ( i \omega  ) }{\partial ( i \omega) }   = 
\frac{ \partial \Sigma_{\sigma} ( i \omega  ) }{\partial \mu}  
-\sum_{\sigma^{\prime}} \rho_{\sigma^{\prime}} ( 0 )
\Gamma_{\sigma , \sigma^{\prime}} ( i \omega , 0).
 \label{eq:sigmamuflat}
\end{equation}

Next, let us choose in our exact  renormalization group flow equation 
 (\ref{eq:selfRGflow})  
the magnetic field as a flow parameter ($\Lambda =h$).  The relevant
single-scale propagator is then
\begin{equation}
\dot{G}_{\sigma } ( i \omega )  =  - 
 {G}^2_{\sigma } ( i \omega ) \left[ \sigma -  \frac{ \partial 
\Delta_{\sigma} ( i \omega )}{\partial h} \right]      ,
 \end{equation}
and hence
 \begin{eqnarray}
 \frac{\partial \Sigma_{\sigma } (i \omega )}{ \partial h } 
 &  = & 
 - \int_{\omega^{\prime}}  \sum_{\sigma^{\prime}} \sigma^{\prime}
\left[ 1 - \sigma^{\prime} \frac{ \partial 
\Delta_{\sigma^{\prime}} ( i \omega^{\prime} )}{\partial h}   
 \right] G_{\sigma^{\prime}}^2 ( i \omega^{\prime} ) 
 \nonumber
 \\
 &  & \times
\Gamma_{\sigma , \sigma^{\prime}} ( i \omega , i \omega^{\prime}).
 \label{eq:hflowfinal}
 \end{eqnarray}
Noting that
\begin{equation}
 \sigma \frac{\Delta_{\sigma} ( i \omega )}{\partial h }  = \frac{\Delta_{\sigma} ( i \omega )}{\partial \mu }  ,
\end{equation}
Eq.~(\ref{eq:hflowfinal}) can also be written as
 \begin{eqnarray}
 \frac{\partial \Sigma_{\sigma } (i \omega )}{ \partial h } 
 &  = & 
 - \int_{\omega^{\prime}}  \sum_{\sigma^{\prime}} \sigma^{\prime}
\left[ 1 - \frac{ \partial 
\Delta_{\sigma^{\prime}} ( i \omega^{\prime} )}{\partial ( i \omega^{\prime} )}   
- 2 \Delta \delta ( \omega^{\prime} )
 \right] 
 \nonumber
 \\
 &  & \times G_{\sigma^{\prime}}^2 ( i \omega^{\prime} ) 
\Gamma_{\sigma , \sigma^{\prime}} ( i \omega , i \omega^{\prime})
 \nonumber
 \\
 & &
  \hspace{-20mm} = 
 - \int_{\omega^{\prime}}  \sum_{\sigma^{\prime}} \sigma^{\prime}
\left[ 1 - \frac{ \partial 
\Delta_{\sigma^{\prime}} ( i \omega^{\prime} )}{\partial ( i \omega^{\prime} )}   
 \right]  G_{\sigma^{\prime}}^2 ( i \omega^{\prime} ) 
\Gamma_{\sigma , \sigma^{\prime}} ( i \omega , i \omega^{\prime})
 \nonumber
 \\
 &  & 
+ \sum_{\sigma^{\prime}} \sigma^{\prime}   \rho_{\sigma^{\prime}} ( 0 )
\Gamma_{\sigma , \sigma^{\prime}} ( i \omega , 0).
 \label{eq:hflowfinal2}
 \end{eqnarray}
Comparing this with the $U(1)$ spin Ward identity (\ref{eq:WardselfU1spin}),
we conclude that
\begin{eqnarray}  
 \frac{ \partial \Sigma_{\sigma} ( i \omega  ) }{\partial ( i \omega) } &  = &
\sigma \frac{ \partial \Sigma_{\sigma} ( i \omega  ) }{\partial h}  
 -  \sigma \sum_{\sigma^{\prime}} \sigma^{\prime} \rho_{\sigma^{\prime}} ( 0 )
\Gamma_{\sigma , \sigma^{\prime}} ( i \omega , 0).
 \nonumber
 \\
 & & 
 \label{eq:sigmahflat}
\end{eqnarray}
Of particular interest are the above relations for $\omega \rightarrow 0$,
because in this limit the $\omega$-derivative determines the 
wave-function renormalization factor $Z_{\sigma}$ via
 \begin{equation}
 \left. \frac{ \partial \Sigma_{\sigma} ( i \omega  ) }{\partial ( i \omega) }
  \right|_{ \omega =0} = 1 - Z_\sigma^{-1}.
 \label{eq:Zdef}
 \end{equation}
Taking the limit $\omega \rightarrow 0$
in Eqs.~(\ref{eq:sigmamuflat}) and (\ref{eq:sigmahflat}) and
using the fact that due to antisymmetry the effective interaction
at vanishing frequencies has the form
 \begin{equation}
U^{(4)}_{ \sigma , \sigma^{\prime} }
( 0  , 0 ; 0  , 0  )
 = \Gamma_{ \sigma , \sigma^{\prime}} ( 0,0) =
\delta_{ \sigma , - \sigma^{\prime}} \Gamma_{\sigma , - \sigma},
 \end{equation}
we obtain
 \begin{eqnarray}
\left. \frac{ \partial \Sigma_{\sigma} ( i \omega  ) }{\partial ( i \omega) }
  \right|_{ \omega =0}
& = &  \frac{ \partial \Sigma_{\sigma} ( i 0  ) }{\partial \mu }  -
 \rho_{ - \sigma} (0) \Gamma_{\sigma , - \sigma},
 \\
\left. \frac{ \partial \Sigma_{\sigma} ( i \omega  ) }{\partial ( i \omega) }
  \right|_{ \omega =0}
& = &  \sigma \frac{ \partial \Sigma_{\sigma} ( i 0  ) }{\partial h }  +
 \rho_{ - \sigma} (0) \Gamma_{\sigma, - \sigma}.
 \end{eqnarray}
Adding and subtracting these equations we obtain
 \begin{eqnarray}
2 \left. \frac{ \partial \Sigma_{\sigma} ( i \omega  ) }{\partial ( i \omega) }
  \right|_{ \omega =0}
& = &  \frac{ \partial \Sigma_{\sigma} ( i 0  ) }{\partial \mu } + 
\sigma \frac{ \partial \Sigma_{\sigma} ( i 0  ) }{\partial h } ,
 \label{eq:sigmamuh}
 \\
2 \rho_{ - \sigma} (0) \Gamma_{\sigma , - \sigma} 
& = &   \frac{ \partial \Sigma_{\sigma} ( i 0  ) }{\partial \mu } 
 - \sigma \frac{ \partial \Sigma_{\sigma} ( i 0  ) }{\partial h } .
 \label{eq:rhomuh}
 \end{eqnarray}
Eq.~(\ref{eq:sigmamuh}) is a corrected version of Eq.~(5.80) of 
Ref.[\onlinecite{Hewson93}].

\subsection{Yamada-Yosida identities}

To make contact with the work of
Yamada and Yosida~\cite{Yamada75},
we now assume a flat density of states and
take the limit of an infinite width 
of the conduction electron band,
$D \rightarrow \infty$. Then there is
a simple exact relation between the
occupation  numbers $n_{\sigma}$ 
of the impurity level and the self-energies $\Sigma_{\sigma} ( i0)$ at
vanishing frequency, \cite{Hewson93} 
 \begin{equation}
 n_{\sigma} = \frac{1}{2} - \frac{1}{\pi} \arctan \left[
 \frac{ E_d - \mu - \sigma h + \Sigma_{\sigma} ( i0 ) }{\Delta} \right].
 \label{eq:Friedel}
 \end{equation}
Taking derivatives of this expression with respect to $\mu$ and $h$ we obtain
 \begin{eqnarray}
 \frac{ \partial n_{\sigma} }{\partial \mu } & = & \rho_{\sigma} ( 0 ) \left[
 1 - \frac{ \partial \Sigma_{\sigma} ( i0) }{\partial \mu } \right] ,
 \label{eq:dmu}
 \\
\sigma \frac{ \partial n_{\sigma} }{\partial h } & = &  \rho_{\sigma} ( 0 ) \left[
 1 - \sigma  \frac{ \partial \Sigma_{\sigma} ( i0) }{\partial h } \right] .
\label{eq:dh}
\end{eqnarray}
Hence,
 \begin{eqnarray}
\frac{ \partial \Sigma_{\sigma} ( i0) }{\partial \mu } & = & 1 -
 \frac{1}{  \rho_{\sigma} ( 0 ) }
 \frac{ \partial n_{\sigma} }{\partial \mu } ,
 \label{eq:dsigmamu}
 \\
\sigma  \frac{ \partial \Sigma_{\sigma} ( i0) }{\partial h } & = & 1 - 
 \frac{1}{ \rho_{\sigma} ( 0 ) }
\sigma \frac{ \partial n_{\sigma} }{\partial h } .
\label{eq:dsigmah}
\end{eqnarray}
Substituting these relations into the
Ward identities (\ref{eq:sigmamuh}) and (\ref{eq:rhomuh}) we obtain
 for the $\omega$-derivative of the self-energy
 \begin{equation}
\left. \frac{ \partial \Sigma_{\sigma} ( i \omega  ) }{\partial ( i \omega )}
 \right|_{ \omega =0}   
= 1 - \frac{1}{2 \rho_{\sigma} (0)} \left[ 
 \frac{ \partial n_{\sigma} }{\partial \mu } +
 \sigma
 \frac{ \partial n_{\sigma} }{\partial h } \right],
 \label{eq:omegaderiv1}
  \end{equation}
and for the effective interaction at zero energy
 \begin{equation}
 \Gamma_{\sigma , - \sigma} =  - \frac{1}{2 \rho_{\uparrow} (0) \rho_{ \downarrow} (0)}
 \left[
 \frac{ \partial n_{\sigma} }{\partial \mu } - \sigma
 \frac{ \partial n_{\sigma} }{\partial h } \right].
 \label{eq:vertWard}
 \end{equation}
For $h \rightarrow 0$ the self-energy and the density of states
are independent of the spin projection, so that we may write
$\Sigma_{\sigma} ( i \omega ) = \Sigma ( i \omega )$,
$\rho_{\sigma} (0) = \rho (0)$, and
$\Gamma_{ \sigma , - \sigma} = \Gamma_{\bot}$.
As in Ref.~[\onlinecite{Yamada75}],
 we now focus on the particle-hole symmetric case,
where $ \rho (0) = 1 /( \pi \Delta )$ in the flat band infinite bandwidth limit considered here.
Averaging both sides of Eqs.~(\ref{eq:omegaderiv1}) and (\ref{eq:vertWard})
over both spin projections, we obtain the Ward identities
\begin{eqnarray}
\left. \frac{ \partial \Sigma_{\sigma} ( i \omega  ) }{\partial ( i \omega )}
 \right|_{ \omega =0}   
& = & 1 - \frac{ \tilde{\chi}_c + \tilde{\chi}_s}{2},
 \label{eq:WardZ} 
 \\
 \rho (0)   \Gamma_\bot & = & -  \frac{  \tilde{\chi}_c - \tilde{\chi}_s}{2   },
 \label{eq:Wardint}
\end{eqnarray}
where the dimensionless particle number (charge) and spin susceptibilities are defined by
 \begin{eqnarray}
 \tilde{\chi}_c & = &  \frac{\chi_c}{\rho (0)}  = \frac{\pi \Delta}{2} \sum_{\sigma} \frac{ \partial n_{\sigma}}{\partial \mu },
 \label{eq:chic}
 \\
\tilde{\chi}_s & = & \frac{ \chi_s}{\rho(0)} = \frac{\pi \Delta}{2} \sum_{\sigma} \sigma \frac{ \partial 
 n_{\sigma}}{\partial h }.
 \label{eq:chis}
 \end{eqnarray}
The identities (\ref{eq:WardZ}) and (\ref{eq:Wardint}) 
have first been obtained 
by Yamada and Yosida,\cite{Yamada75}
who showed that both sides of these equations
have identical series expansions in powers of $U/ \Delta$.
Such a perturbative proof relies on the assumption that
there are no non-analytic terms which are missed by the series expansions.
Our proof of  Eqs. (\ref{eq:WardZ}) and (\ref{eq:Wardint}) 
given above shows more clearly that these 
identities are a direct consequence of the
$U(1) \times U(1)$ symmetry 
associated with particle number and spin conservation of the AIM in a magnetic field.
Moreover, our derivation of  Eqs. (\ref{eq:WardZ}) and (\ref{eq:Wardint})  
is non-perturbative, because it relies only on the  symmetries of the
AIM and the associated functional
Ward identities, and on
the exact renormalization group flow equation for the irreducible self-energy.

Using relation (\ref{eq:Zdef}), we see that
the identity  (\ref{eq:WardZ}) implies that the wave-function renormalization
factor of the AIM can be expressed in terms of  the  susceptibilities as
 \begin{equation}
 Z = \frac{2}{ \tilde{\chi}_c + \tilde{\chi}_s }.
 \end{equation}
The other Ward identity (\ref{eq:Wardint}) allows us to express the
imaginary part of the retarded self-energy to the susceptibilities.
Therefore we recall that the skeleton equation for the self-energy
of the AIM (which is a consequence of the Dyson-Schwinger equation, which in 
turn follows within a functional integral approach from the invariance
of the functional integral under infinitesimal shifts of the 
fields\cite{Kopietz10,ZinnJustin02})
implies the following exact expression 
for the imaginary part of the self-energy of the symmetric AIM in the
infinite bandwidth limit,\cite{Hewson93}
 \begin{equation}
 {\rm Im}\, \Sigma ( \omega + i 0 ) =
  -   \left( \frac{\Gamma_\bot}{\pi \Delta } \right)^2   \frac{\omega^2}{2 \Delta} 
+ {\cal{O}} ( \omega^3 ).
 \label{eq:sigmaexpandIm2}
\end{equation}
Substituting the identity (\ref{eq:Wardint})
for the dimensionless effective interaction
$ \Gamma_{\bot} /(\pi \Delta ) = \rho (0) \Gamma_{\bot}$, we obtain
\begin{equation}
 {\rm Im}\, \Sigma ( \omega + i 0 ) =
  -   \left( \frac{ \tilde{\chi}_c - \tilde{\chi}_s}{2} \right)^2  \frac{\omega^2}{2 \Delta} 
+ {\cal{O}} ( \omega^3 ).
 \label{eq:sigmaexpandIm}
\end{equation}
On the imaginary frequency axis, the low-frequency expansion of the
self-energy of the symmetric AIM in the infinite bandwidth limit is therefore
\begin{eqnarray}
  \Sigma ( i \omega  )  & =  & \frac{U}{2} + 
\left( 1- \frac{\tilde{\chi}_c + \tilde{\chi}_s}{2} \right)  i \omega
+ i  \frac{(\tilde{\chi}_c - \tilde{\chi}_s)^2}{8 \Delta} \omega^2 {\rm sgn}\, \omega
 \nonumber
 \\
 &+  & 
  \mbox{analytic terms ${\cal{O}} (\omega^2)$}.
 \label{eq:sigmaexpandMat}
\end{eqnarray}
Taking the different notations for the dimensionless  susceptibilities into 
account,\cite{footnotenorm}
Eq.~(\ref{eq:sigmaexpandMat}) agrees with the expressions derived
by Yamada and Yosida.\cite{Yamada75}

\section{Conclusions}

In this work we have used modern functional
methods to give a non-perturbative proof of the
Yamada-Yosida identities, which express the coefficients 
in the low-frequency expansion of the self-energy of 
the Anderson impurity model in terms of thermodynamic 
susceptibilities.  In contrast to the derivation of these relations
given by Yamada and Yosida,\cite{Yamada75} which is based on a series expansion
in powers of the interaction, our non-perturbative proof
relies on exact Ward identities and on an exact
renormalization group  flow equation for the irreducible self-energy.
From our derivation it is obvious that
the Yamada-Yosida identities are a direct
consequence of the $U(1) \times U(1)$ symmetry of the Anderson impurity model
associated with the conservation of the
particle number and the total
spin component in the direction of an external magnetic field.

We have also presented a thorough derivation of the 
general functional Ward identities of the AIM
due to particle number and spin conservation.
Furthermore, we have shown that various identities relating
the derivatives of the self-energy with respect to frequency, chemical potential, and
magnetic field can be obtained by combining the Ward identities 
with exact renormalization
group flow equations for the self-energy.

\section*{ACKNOWLEDGMENTS}
We thank S. Andergassen, V. Meden, and M. Pletyukhov for  discussions.
This work was financially supported by
FOR 723 and by the DAAD/CAPES PROBRAL-program.


\begin{thebibliography}{99}
\bibitem{footnoteMorris} 
Alternatively, we arrive at the Ward identity (\ref{eq:WardU1infinite})
without  invoking the Morris-Lemma if we 
choose the bosonic frequency $\bar \omega$ in Eq.~(\ref{eq:Wardparticle}) 
to be equal to $2 \pi/\beta$, replace the integral $\int_{\omega^{\prime}}$ by the sum $\frac{1}{\beta}\sum_{\omega_n^{\prime}}$ with $\omega_n^{\prime} = (2n+1)\pi/\beta$, and finally take the limit $\beta \to \infty$.
\bibitem{footnotenorm}
Our dimensionless particle number (charge) and spin susceptibilities 
$\tilde{\chi}_c$ and $\tilde{\chi}_s$
are related to the quantities 
$\tilde{\chi}_{\rm even} = \tilde{\chi}_{ \uparrow \uparrow}$ and
$\tilde{\chi}_{\rm odd} = \tilde{\chi}_{ \uparrow \downarrow}$ 
introduced by Yamada and Yosida~\cite{Yamada75} and by
Oguri \cite{Oguri01} via
$ \tilde{\chi}_{c} =  \tilde{\chi}_{ \uparrow \uparrow} +  
\tilde{\chi}_{ \uparrow \downarrow}$ and
$ \tilde{\chi}_{s} =  \tilde{\chi}_{ \uparrow \uparrow} -
\tilde{\chi}_{ \uparrow \downarrow}$, so that
$(\tilde{\chi}_c + \tilde{\chi}_s) /2 = \tilde{\chi}_{\rm even} =
\tilde{\chi}_{ \uparrow \uparrow} $ and
$(\tilde{\chi}_c - \tilde{\chi}_s) /2 = \tilde{\chi}_{\rm odd} =
\tilde{\chi}_{ \uparrow \downarrow} $,
\end{thebibliography}
\end{document}